# Experimental measurement of radiation dose in a dedicated breast CT system*

SHEN Shan-Wei (申善威)[1,2,3;1)]  WANG Yan-Fang(王燕芳)[2,3]  SHU Hang(舒航)[2,3]  TANG Xiao (唐晓)[2,3]
WEI Cun-Feng (魏存峰)[2,3] SONG Yu-Shou(宋玉收)[1]  SHI Rong-Jian (史戎坚)[2,3] WEI Long(魏龙)[2,3]

1 (College of Nuclear Science and Technology, Harbin Engineering University, Harbin 150000, China)

2 (Key Laboratory of Nuclear Analytical Techniques, Institute of High Energy Physics, CAS, Beijing 100049, China)

3 (Beijing Engineering Research Center of Radiographic Techniques and Equipment, Beijing 100049, China)

**Abstract:** Radiation dose is an important performance indicator of a dedicated breast CT (DBCT). In this paper, the method of putting thermoluminescent dosimeters (TLD) into a breast shaped PMMA phantom to study the dose distribution in breasts was improved by using smaller TLDs and a new half-ellipsoid PMMA phantom. Then the weighted CT dose index (CTDIw) was introduced to average glandular assessment in DBCT for the first time and two measurement modes were proposed for different sizes of breasts. The dose deviations caused by using cylindrical phantoms were simulated using the Monte Carlo method and a set of correction factors were calculated. The results of the confirmatory measurement with a cylindrical phantom (11cm/8cm) show that CTDIw gives a relatively conservative overestimate of the average glandular dose comparing to the results of Monte Carlo simulation and TLDs measurement. But with better practicability and stability, the CTDIw is suitable for dose evaluations in daily clinical practice. Both of the TLDs and CTDIw measurements demonstrate that the radiation dose of our DBCT system is lower than conventional two-view mammography.

## 1 Introduction

Breast cancer is the most common malignancy in women, causing the death of hundreds of thousands of women, and the morbidity rate is increasing year by year. The early diagnosis and treatment of breast cancer is important for prognosis, improving the quality of patients' life and reducing the cost of treatment. DBCT overcomes the shortcomings of many other breast examination methods: the X-ray of DBCT doesn't penetrate the chest cavity as in conventional CT so won't produce additional doses; breasts are not compressed in the examination of DBCT and patients may feel more comfortable than mammography; and the three-dimensional images of the breast structure in the natural state overcome overlap of breast tissues in mammography, so images of DBCT can accurately display the locations, shapes, number and sizes of the breast lesions, which is helpful to distinguishing benign breast tumors from malignant ones when it is combined with the observation of other features of tumor , for example, whether it has metastasized; besides, DBCT can guide biopsy for clinicopathologic analysis. For radioactive diagnostic equipments, the radiation dose is an important criterion to evaluate their security, and the potential radiation damage must be strictly controlled. DBCT is no exception, good image quality of which is meaningful only when the radiation dose is in safe range. For any method attempting to improve the image quality such as changing geometry, scanning mode of DBCT system, optimizing experimental parameters and so on, the prerequisite of which is that the radiation dose should not be increased. So methods that can evaluate the radiation dose accurately and objectively are needed in the process of debugging and running of DBCT. At present, Monte Carlo simulations and experiments are the two main ways to study the absorbed dose in breast tissue. Through Monte Carlo simulations, any factors affecting the image quality such as the size, shape and material of phantoms, system geometry and tube voltage and current can be calculated separately [1, 2], but the simulation results need to be verified by experiments. Russo et al examined dose distribution by placing TLDs in a half-ellipsoid polymethyl methacrylate (PMMA) phantom [3], whose results showed that the DBCT delivered a more uniform dose to breasts, so the risk is minor for patients relative to mammography. But the size of TLDs they used is relative large (3mm × 3mm × 0.9mm) to the size of phantom and three TLDs were

Received date

*Supported by National Natural Science Foundation of China (81101045) and Knowledge Innovation Project of Chinese Academy of Sciences (KJCX2-EW-N06)

1) E-mail: shenshanwei@ihep.ac.cn

located each of the six cavities, which is bound to increase the influence on primary dose distribution in phantom. Furthermore, the positions of breasts in examination and the radiation dose in chest wall are not considered during their measurements. The ionization chamber has been used to measure the absolute dose in DBCT. Boone et al used the ionization chamber to measure the air karma at the isocenter of a cylindrical phantom and calculated the average glandular dose by multiplying the normalized glandular dose coefficients for CT (DgNCT) calculated by Monte Carlo simulations [4]. But the shapes of cylindrical phantoms are different from breasts, so dose calculation deviationss may be inevitable.

In this paper, two experimental methods were carried out to study the radiation dose of DBCT. On one hand, we improved the TLDs dose measurement method and smaller size of cylindrical TLDs and a half-ellipsoid phantom with cavities in the breast and chest wall part were used. On the other hand, the standard dose evaluation method in conventional CT, CTDIw, was introduced to the DBCT dose measurement for the first time. The dose differences caused by using cylindrical phantoms, which are different from the half-ellipsoid shape of the breast, were calculated by Monte Carlo simulations and correction factors of CTDIw were given corresponding to different breast sizes to avoid underestimating the real dose in breasts.

## 2 Materials and methods
### 2.1 DBCT system

The X-ray tube used in our DBCT system operates between 5kV and 75 kV with a current range of 0-17.5 mA, which has a tungsten anode with a minimum focal spot size of 1.0mm and an inherent filtration of 0.8mm Be. The flat panel detector used in DBCT is the Varian PaxScan 2520D/CL, the size and resolution of which are suitable to breast imaging. The X-ray tube and detector were coupled to the slip ring at a certain relative position to constitute the main structure of DBCT system. Table 1 shows the operation conditions of DBCT determined in early work under which good images quality can be obtained.

Table 1. DBCT geometry and operation parameters

| Source angle α | About 15 ° |
|---|---|
| SOD | 60cm |
| Tube voltage | 70kV |
| Tube current | 8mA |
| Additive filter | 8mm Al |
| Exposure time | 15s |
| Projections | 450 |

### 2.2 Dosimeters and phantoms

A group of 36 TLDs (LiF: Mg, Cu, P, Φ1.5mm×0.8mm) as shown in Fig.1 (a) were used for dose distribution measurements. The breast phantom still has a shape of half-ellipsoid composed by two halves of block machined from one PMMA cylinder of 14cm diameter (Fig. 1(b)), which has a 8cm half-ellipsoid breast part, a 3cm cylindrical chest wall part and a 2.5cm auxiliary suspension structure. A total of 18 TLDs were placed in the breast part and 15 TLDs in the chest wall part, so dose of these two parts can be measured at the same time.

The phantom used for CTDIw measuring experiments has a cylindrical shape and PMMA material similar to the standard CTDI phantom used in the conventional CT as shown in Fig.1(c) , but has a length of 13cm and a diameter of 11cm so that the result can be compared with TLDs experiment. The CTDIw phantom has a hole in the isocenter and 8 holes at the periphery where ionization chambers can be put, and there were also 9 PMMA sticks can be used to fill the rest holes during measurement. The length of the ionization chamber used here is a standard 10cm (PTW, Freiburg, Germany), which can perform the dose integration process in measurements and get values in mGy cm to be used for CTDI calculation.

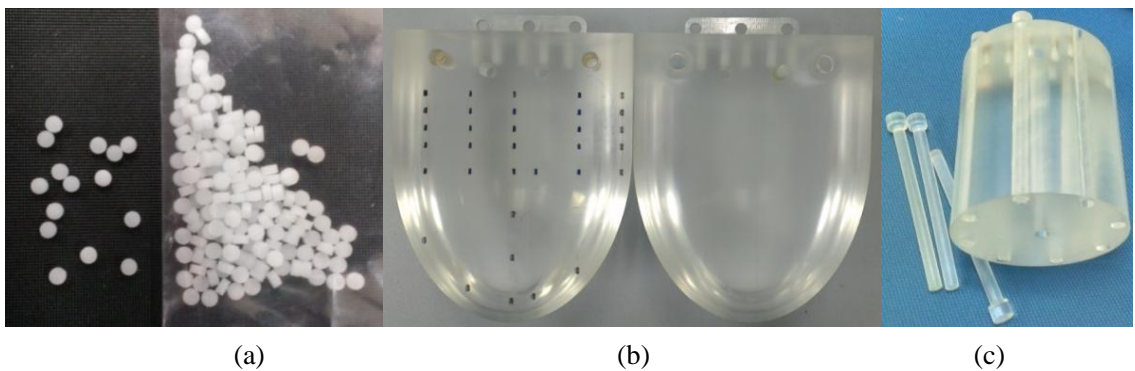

(a)          (b)          (c)

Fig. 1. (a) TLDs used in dose distribution measurement. (b) The half-ellipsoid phantom used in TLDs measurement experiment. (c) The cylindrical phantom used in CTDIw measurement experiment.

### 2.3 TLDs dose measurements

TLDs are commonly used to measure the absorbed dose inside phantoms in dosimetry measurements because they can be made into different shapes and sizes. In this paper, we followed the standardized processing procedures given in [5] and selected 36 pieces with the best homogeneity and reproducibility from 300 TLDs for dose measuring. Then, the monoenergetic gamma beam (Cs-137) was used to calibrate the TLDs because the radiation dose in every exposure can be controlled easily and accurately. TLDs were exposed under the dose of 1mGy, 6mGy, 10mGy, 15mGy and 20mGy respectively in calibrations, and 24 hours later the 36 TLDs were read and annealed after each exposure. But the energy response differences of TLDs must be considered here because different effective energies of beams were used in calibration (622keV) and measurement (40keV). According to the energy response curve given by manufacturer, luminous efficiency of LiF: Mg, Cu, P at 40keV is about 1.5 times higher than at 622keV, so all the readouts in calibrations should be given a correction factor of 1.5 when plotting the relation curve at 40keV. The overall relative standard deviation of the 36 TLDs is about 10.0%, including the errors due to dosimeters screening and calibration. In this way, once we get the readouts of TLDs after exposure, the absorbed dose can be obtained with linear interpolation method base on the relation curve as shown in Fig. 2.

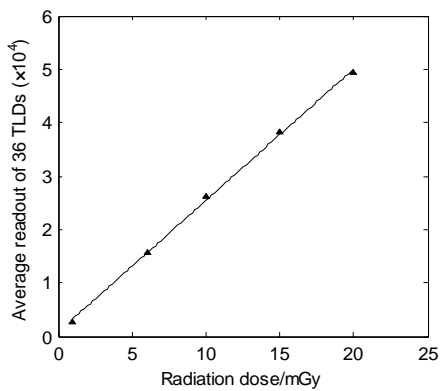

Fig. 2. TLDs dose response curve (effective energy of X-ray beam is 40 keV).

In measurement, the phantom must be located at the right position to ensure the X-ray beam only irradiate the breast part of the half-ellipsoid phantom and the chest wall part is out of FOV of DBCT as in clinical practice. In that case, the dose in breast part is high and good image quality of breast can be obtained, but the dose of chest wall part is relatively low and the unnecessary dose of which can be avoided at the same time.

### 2.4 CTDIw measurements and corrections

CTDI measurements with the pencil chamber and two kinds of cylindrical phantoms, 16cm diameter for head and 32cm diameter for trunk, are the standard methods currently used in the dose assessment of conventional CT, which was proposed by Shapo et al in 1981 for the first time [6], and has been adopted and defined by FDA, IEC, CEC, IAEA and other organizations. Leitz et al introduced a practical approach for measuring the average absorbed doses in CTDI PMMA phantoms and effective doses to the patients combining the tissue weighting factors of different parts of the body in 1995, assuming there is a linear decrease in dose between the periphery and the centre of the phantom [7]. In this method, five CTDI measurements were taken, one in the centre and four in the periphery of the CTDI phantom, then this five results were used to yield one CTDI value with the weighting factor of 1/3 for the centre CTDI and 2/3 for the averaged peripheral CTDI respectively, which was unified defined as the weighted CTDI (CTDIw) later. Comparison with the dose evaluations based on Monte Carlo simulations confirms the validity of this method. For a beam width W less than the length of the chamber L (10cm), CTDIw is given by the empirical equation [8]:

$$CTDI_W = (\frac{1}{3} Dcentre + \frac{2}{3} Dperiphery)L/W, \quad (1)$$

Where $D_{centre}$ is the dose measured in the centre of the CTDI phantom and $D_{periphery}$ is the average of the doses measured at the outer symmetrical four chamber positions of the phantom.

When the beam width W is greater than the length of the chamber L, W get the value of L, CTDIw is given by the empirical equation:

$$CTDI_W = \frac{1}{3} Dcentre + \frac{2}{3} Dperiphery. \quad (2)$$

CTDIw was also used in dose assessment of cone beam CT system (CBCT) [9]. Amer et al believed that although CBCT is not a sequential, slice based technique, CTDI is impractical for measuring dose in CBCT, the standard 10cm chamber can continued be used to give a reasonable estimate of the dose in a certain region of CBCT FOV and the empirical equation (1) and (2) still can be used to calculate the CTDIw (CBDIw for CBCT). DBCT is a cone beam CT which images the breast in hundreds of directions in 360°, so this method is also suitable for DBCT dose measurement.

Corresponding to different sizes of breasts, two ways of placing the phantom and the ionization chamber when measuring CTDIw in DBCT system were used as shown in Fig. 3.

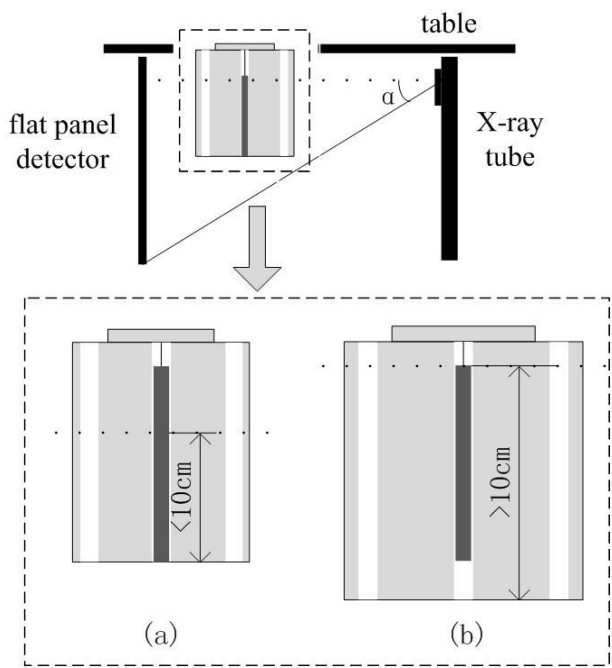

Fig. 3. The ways of placing the ionization chamber in the phantoms (central positions) when measuring CTDIw in DBCT system. (a): the exposed length of phantom is less than 10cm. (b): the exposed length of phantom is greater than 10cm.

Due to the special half ellipsoid shape of the breast the dose estimates is consistently lower with cylindrical phantoms [2], so CTDIw should be corrected to maximize the accuracy of the measurements. In order to get the correction factors, dose simulation with half ellipsoid and cylindrical PMMA phantoms of different lengths (L) and diameters (D) were made respectively to give two groups of average glandular dose, then the dose ratios of the two groups were calculated to be used as the correction factors of CTDIw. In the simulation with half ellipsoid phantoms, we followed the way of constructing breast phantoms and used the X-ray spectrum (70kV, 8mmAL filter) given in [10]. The results of two groups of simulation and the dose ratios are shown in Table 2.

Although the simulation results of average glandular dose with cylindrical phantoms is only a little lower than that with half ellipsoid phantoms as listed in table 2, the corrected CTDIw by CFs can minimize the deviation of measurements to avoid dose underestimates in the process of debugging and running of the DBCT.

Experiment was performed with a cylindrical breast phantom to verify the practicability of CTDIw in dose evaluation of the DBCT system. The phantom was placed in FOV of DBCT as shown in Fig. 3(a), and the CTDIw was calculated by empirical equation (1) with the W=8cm and L=10cm.

## 3 Results
### 3.1 Dose distribution in breast and chest wall

TLDs were read 24 hours later after the half ellipsoid phantom was exposed, and readouts were converted into dose by linear interpolation according tothe TLDs dose response curve. Values obtained in the 33 positions are shown in Fig. 4. The rapid decreasing of dose values between the breast and the chest wall confirmed the phantom was placed at the right place as described above. In the breast part, dose in both of the directions of the breast, radial and longitudinal, have a gradual increment, which is not exactly the same as the results of Russo et al, because of the consideration of phantom position in our measurement, but a more uniform dose distribution inside the breast obtained in DBCT examinations compared with that obtained in traditional mammography was observed in both of measurements. So if the same dose was delivered to breasts in one examination of mammography and DBCT, a more uniform dose distribution of the latter inside the breasts will be safer obviously.

Table 2. The simulation results of average glandular dose with two shapes of breast phantoms.

| average glandular dose Gy per million photons($\times 10^{-5}$) | | phantoms | | dose ratios (correction factors, CFs) |
|---|---|---|---|---|
| | | cylindrical | half-ellipsoid | |
| breast sizes (D/L) | 9cm/6cm | 2.47 | 2.60 | 1.05 |
| | 10cm/7cm | 2.44 | 2.48 | 1.02 |
| | 11cm/8cm | 2.37 | 2.41 | 1.02 |
| | 12cm/9cm | 2.35 | 2.40 | 1.02 |
| | 13cm/10cm | 2.32 | 2.38 | 1.03 |
| | 14cm/11cm | 2.22 | 2.36 | 1.06 |
| | 15cm/12cm | 2.06 | 2.29 | 1.11 |

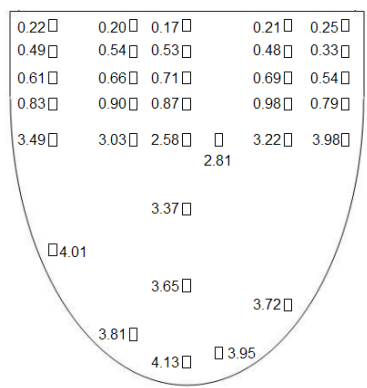

Fig 4. Doses obtained by TLDs in the 33 positions of the half-ellipsoid phantom (mGy).

In order to give an estimation of the average glandular dose, the values of 13 TLDs in breast part of the phantom were averaged, and the result, 3.45mGy (the overall error is also 10%), can be regarded as a rough estimate of the average glandular dose in DBCT examination because of the low accuracy of TLDs in dose measurement. Even so, the average glandular dose obtained by TLDs can be used as a reference value for other dose measurement methods, for example the CTDIw used in this paper and the simulation methods. After the pencil ionization chamber was placed at the predetermined position as shown in Fig.3 (a), dose were read three times for each position and the average values were calculated to reduce the measuring deviations. Measurement results and the CTDIw calculated by empirical equation (1) were listed in Table 3. Then the CTDIw multiplied by the correction factors (CFs) corresponding to the size of breast (11cm/8cm) to get the estimation of average glandular dose.

### 3.2 simulations of average glandular dose

To further evaluate the validity of CTDIw, we compared it with the average glandular dose simulation result made by Tang et al using GATE (Geant 4 application for tomographic emission) [10]. In their simulation of DBCT, half ellipsoid breast phantom with the material of 50% glandular and 50% adipose tissue covered by 3-5mm skin were used , which is similar to the real breast. Besides, monoenergetic X-ray beams were used in simulations to optimize the spectrum, the density different of glandular and adipose was also considered so the results were supposed to be credible and have a good reference value. According to the results obtained in [10], when the experiment conditions is the same as we used in this paper, for a breast of 11cm/8cm size, the average glandular dose per mAs is 2.95E-05Gy. So for a single scan of DBCT the overall dose is about 3.54mGy obtained by 2.95E-05Gy multiplied by 120mAs (15s, 8mA). Besides, the different material must be taken into account when the dose results of experimental study and the simulation studies were compared with each other. According to the estimation in [3] of the discrepancies in absorbed dose due to the different materials (PMMA and 50-50 breast tissue), the value of ratio DPMMA with respect to D50-50 is about0.9 in the range of effective energies from 35.7 to 44.4 keV (which is 40keV for our beam). So we consider that the average glandular dose obtained by using PMMA in our measurement is about 10% less than that obtained byusing 50-50 breast tissue in simulations. The converted results of the two experimental studies and the simulation study by Tang et al are listed in Table 4.

Table 3. The results of CTDI100 and CTDIw

| phantom (D/L) | breast (D/L) | $CTDI_{100,periphery}$ | $\overline{CTDI_{100,periphery}}$ | $CTDI_{100,cenrte}$ | $CTDI_w$ | $CF \times CTDI_w$ |
|---|---|---|---|---|---|---|
| 11cm/13cm | 11cm/8cm | 31.2mGy•cm<br>30.5mGy•cm<br>31.0mGy•cm<br>30.2mGy•cm | 32.0mGy cm | 26.3mGy cm | 3.65mGy | 3.72mGy |

Table 4. The average glandular dose (AVG) of DBCT obtained by three methods (the breast size is 11cm/8cm)

| AVG/mGy | Gate simulation (Tang et al.) | TLDs | $CTDI_w$ |
|---|---|---|---|
| | 3.54(1) | 3.84(1.08) | 4.13(1.17) |

When comparing the three estimations of the average glandular dose, the agreement of two experimental results is found to be satisfactory generally when taking the measurement errors of TLDs into account. But both of the experimental measurements get higher results than the simulation, especially the CTDIw, which is about 17% higher than Gate results. There are many reasons for causing these errors. For simulation studies, estimation errors of the actual operating conditions of DBCT is inevitable, such as the X-ray spectrum, the irradiation flux of photons, the system geometry, the phantoms material and so on. Because the construction of simulation DBCT system tend to be idealized, so the result of which must be validated by experimental results. Besides, owing to the variability of TLDs, readouts of them may not be the same each time even after careful screening and accurate calibration. And after many times of irradiation and annealing, the TLDs became insensitive, which will cause small readouts and dose values. Then for the method of CTDIw proposed in this paper, there are also many factors can lead to errors. First, the empirical formula (1) and (2) is based on the assumption that the dose has a linear decrease in the radial direction of the cylindrical phantom, while the actual situation may not be like that. In addition, even after correction, the dose measurements using a cylindrical instead of half ellipsoidal phantom may cause errors because different shapes may cause different dose distribution in the phantoms. Finally, the effectiveness of CTDIw has always been controversial, because of using different range of integration and length of ionization chambers will get different results, and the most appropriate combinations have not been determined yet for different CT systems. In this paper, we used a 10cm ionization chamber and single phantom to measure CTDIw. So for the breast length less than 10cm, a part of the ionization chamber (2cm for the breast of 11cm/8cm size) is out of the FOV, where the dose we supposed to be zero. But as can be seen from Fig. 4, the chest wall part also has dose deposit because of the X-ray scattering, which will make CTDIw results calculated by the empirical formula (1) higher than virtually dose, and that is also the main reason why CTDIw get the highest result in three methods. In contrast, when the breast length is greater than 10cm, the CTDIw value is closely related to the position of ionization chamber in phantoms during measurement. Because for the FOV of our DBCT system using a half cone beam, radiation dose reduces from top to bottom on longitudinal [9], which means the CTDIw obtained in the upper 10 cm region of the FOV as shown in Fig. 3(b) is higher than that obtained across the whole exposed region of phantom. Therefore, in the the R & D process of DBCT, variety of methods should be used to study the radiation dose to ensure the accuracy of dose evaluation. However, once the DBCT system access to clinical trials or practical application, the CTDIw can be adopted as a standard method like in conventional CT for the assessment of average glandular dose in view of its usability and good stability.

At present, conventional mammography is still the "Golden Standard" in the breast cancer diagnosis, and which has the specific limit of average glandular dose in examination. In the USA the guidelines of limitations to the maximum mean dose to the radiosensitive glandular tissue (MGD) delivered by a single view suggested by American College of Radiology (ACR) is 3mGy for a 4.2cm thick compressed breast, consisting of 50% glandular and 50% adipose tissue, either for full field digital mammography or screen-film mammography[11]. Hence, DBCT can assume the average glandular dose of a two-view exam, 6mGy, as a reference limiting value for DBCT. In Europe, this reference limiting value of MGD is set as 5mGy for a two-view exam in mammography for an average compressed breast of 5.3cm [12]. To compare the DBCT with mammography on an equal-dose basis, the MGD to the single breast in DBCT imaging should be not higher than that (5-6mGy). As can be seen from the Table 4, either for U.S. or European standards, our DBCT system is safe under the current conditions. In Addition, the results obtained by TLDs indicate the dose distribution of DBCT is more uniform than mammography. The results in [13] obtained by Boone et al show that the parts of the breast where the X-ray beam penetrates can be several times the absorbed dose of the parts on the opposite side in mammography, so even the two systems give the same dose to patients, DBCT still is the safer one.

In this work, the radiation dose of our DBCT system was evaluated by experimental methods. The smaller TLDs and a new half ellipsoidal phantom with more cavities inside were used to measure the dose distribution in the breast and chest wall. Besides, the phantoms were set at a fixed position in measurement just like in clinical practice because different positions in FOV may affect the dose distribution in phantoms. The results reconfirmed that DBCT delivered a more uniform dose distribution than mammography. Finally, an estimation of the average glandular dose was obtained by averaging the corrected values of TLDs by dose response curve of the breast part. On the other hand, for the first time we proposed to use the concept of CTDIw combining with a 10cm ionization chamber to evaluate the radiation dose of the dedicated breast CT systems, and two measurement modes were used for different size of breasts. A group of correction factors related to the different shapes of phantoms were calculated by Monte Carlo simulations for correcting the CTDIw to get the average glandular dose. Comparison with TLDs and Gate simulation results show that the CTDIw gives a useful, relatively conservative overestimate of the average glandular dose, but whose practicability and stability is better so it is suitable for dose assessment in clinical practice.

Comparison with the dose limits of mammography shows that our DBCT system delivered a lower dose to patients when it obtains high quality 3-D images, that is to say there is still much potential room for DBCT to improve the image quality within the dose limits because a higher dose can bring better signal-to-noise ratio theoretically.

In future studies, we should continue to work on improving the image quality of DBCT by changing the experiment conditions and reducing the patient dose at the same time. Furthermore, we will give CTDIw more accurate correction factors to assess the real dose of DBCT by analyzing the comprehensive factors that are influential in measurements of CTDIw in addition to the shape of phantoms, for example the different tube outputs, breast sizes and measurement modes using pencil chamber, so CTDIw can be served as the standard dose assessment method as in conventional CT in the future.